\documentclass[preprint,12pt]{elsarticle}

\journal{Physics of the Dark Universe}

\usepackage{graphicx}
\usepackage{lineno,hyperref}
\modulolinenumbers[5]
\usepackage{array}
\usepackage{microtype}
\usepackage{float}
\usepackage{amsfonts}
\usepackage{mathrsfs}
\usepackage{amsmath}
\usepackage{epstopdf}
\hypersetup{colorlinks,citecolor=blue}
\usepackage{multirow}
\begin{document}
\begin{frontmatter}
\title{Anisotropic cosmology using observational datasets: exploring via machine learning approaches}

\author[addr1]{Vinod Kumar Bhardwaj}\ead{dr.vinodbhardwaj@gmail.com}
\author[addr2]{Manish Kalra}\ead{manishkalra2012@gmail.com}
\author[addr3]{Priyanka Garg}\ead{pri.19aug@gmail.com}
\author[addr4]{Saibal Ray}\ead{saibal.ray@gla.ac.in}

\address[addr1]{Department of Mathematics, GLA University, Mathura 281406, Uttar Pradesh, India}
\address[addr2]{Department of Electronics and Communication Engineering, GLA University, Mathura 281406, Uttar Pradesh, India}
\address[addr3]{Department of Mathematics, GLA University, Mathura 281406, Uttar Pradesh, India}
\address[addr4]{Centre for Cosmology, Astrophysics and Space Science (CCASS), GLA University, Mathura 281406, Uttar Pradesh, India}
		
\date{Received~~2025 May 3;~accepted~~2025~~month day}

\begin{abstract}
In the current study, we present the observational data constraints on the parameters space for an anisotropic cosmological model of Bianchi I type spacetime in general relativity (GR). For the analysis, we consider observational datasets of Cosmic Chronometers (CC), Baryon Acoustic Oscillation (BAO), and Cosmic Microwave Background Radiation (CMBR) peak parameters. The Markov chain Monte Carlo (MCMC) technique is utilized to constrain the best-fit values of the model parameters. For this purpose, we use the publicly available Python code from CosmoMC and have developed the contour plots with different constraint limits. For the joint dataset of CC, BAO, and CMBR, the parameter’s best-fit values for the derived model are estimated as $ H_0 = 69.9\pm 1.4$ km/s/Mpc, $ \Omega_{m0}=0.277^{+0.017}_{-0.015}$, $ \Omega_{\Lambda 0} = 0.722^{+0.015}_{-0.017}$,  and $\Omega_{\sigma 0} = 0.0009\pm0.0001$. To estimate $H(z)$, we explore machine learning (ML) techniques like linear regression,  Artificial Neural Network (ANN), and polynomial regression and thereafter analyze the results with the theoretically developed $H(z)$ for the proposed model. Among these ML techniques, the polynomial regression exceeds the performance compared to other techniques. Further, we also note that larger dataset provides a better understanding of the cosmological scenario in terms of ML view point.\\

\end{abstract}

\end{frontmatter}

\section{Introduction}

Cosmology, in the present days, is the scientific study of the large-scale structure of the universe as a whole. It deals with the origin, evolution, dynamics, and eventual fate of the universe. In the early 20th century, Edwin Hubble observed that galaxies were moving away from us in all directions, suggesting that the universe is expanding \cite{ref1,ref2,ref3,ref4,ref5}. This observation formed the framework for the Big Bang theory, which proposes that the universe began as an extremely dense and hot state approximately 13.8 billion years ago. However, scientists expected that the expansion of the universe would gradually slow down over time due to the gravitational attraction between galaxies. This assumption was based on the influence of the matter, both visible and dark, which was thought to dominate the universe's energy density. Surprisingly, in the late 1990s, observations of the distant supernovae revealed that the expansion of the universe was not slowing down rather accelerating. This unexpected findings revolutionized the conventional concepts of cosmology and led to the proposal of {\it dark energy} as a mysterious force driving this acceleration \cite{ref6,ref7,ref8,ref9}. Essentially, this dark energy is a hypothetical form of energy that permeates space and exerts a negative pressure, causing the expansion of the universe to accelerate. 

To explain the accelerated expansion, cosmologists have explored various theoretical frameworks. One such area of study is to modify the laws of gravity themselves. Modified gravity theories \cite{ref10,ref11,ref12,ref13,ref14,ref15,ref16,ref17,ref18,ref19,ref20} propose alterations to Einstein's equations of gravity, known as the Einstein-Hilbert (EH) action, which govern the curvature of spacetime in the presence of matter and energy. However, the EH action is a fundamental concept in the framework of GR. The action is a mathematical expression that encapsulates the dynamics of gravity, providing the foundation for Einstein's field equations. In modified theories of gravity, modifications are made to the EH action to account for deviations from standard GR \cite{ref21,ref22,ref23,ref24,ref25,ref12,ref27,ref28}. These modifications introduce additional terms or functions into the action, altering the dynamics of gravity on cosmological scales. By modifying the EH action, these theories offer alternative explanations for phenomena such as the accelerated expansion of the universe, without necessarily requiring the existence of exotic dark energy. They provide avenues for exploring the nature of gravity beyond Einstein's GR and testing the boundaries for our understanding of the fundamental physics on the cosmological scales.

Bianchi cosmological models of the universe have gained an important role in observational cosmology in recent years, especially via the WMAP data \cite{ref4,ref29,ref30} modeling that is similar to the Bianchi morphology \cite{ref32,ref33,ref34,ref35,ref36,ref37}. To clarify the relational aspect we would like to mention a bit that in their study on a violation of cosmological isotropy, Jaffe et al. \cite{ref32} have applied WMAP data to explore evidence of vorticity and shear at large angular scales. Similarly, Hoftuft et al. \cite{ref37} employed the five-year WMAP data to show increasing evidence for hemispherical power asymmetry.

Bianchi I metric is a solution to Einstein's field equations of GR, describing a homogeneous but anisotropic universe. This means that while the universe appears uniform at every point (homogeneous), it does not look the same in all directions (anisotropic). This metric is particularly important in cosmology for understanding more general types of universes beyond the simple isotropic models like the Friedmann-Lema\^{i}tre-Robertson-Walker (FLRW) metric. Bianchi I metric is often used to model the early universe, where anisotropies could have played a significant role before isotropization processes smoothed out the differences \cite{ref38,ref39,ref40}. It helps in understanding models of the universe that are not perfectly isotropic, providing insights into possible deviations from the standard cosmological model \cite{ref41,ref42,ref43}. The Bianchi Type I metric is also studied in the context of modified theories of gravity to explore how these theories might influence the evolution and structure of an anisotropic universe. Essentially, modified gravity theories are extensions or alternatives to GR that attempt to address various issues such as dark energy, dark matter, and the cosmological constant problem. 

The ML algorithms are useful for parametric extraction of cosmological models and prediction of the Hubble parameter. Now-a-days researchers have been exploring machine learning (ML) techniques for constraining the parameters of the cosmological models utilizing various observational datasets and to analyze the status of modified gravity theories. ML algorithms basically interpret the data to perform for the required predictions. The datasets act as the fuel for the machine learning algorithm. However, the algorithms do not require any specific theoretical model to analyze and predict rather all the predictions are merely based on the dataset. The ML techniques have been explored by researchers in almost all the domains of human endeavors varying from space to everyday human life. The ML techniques are mainly founded on the principle of making predictions based on data. Artificial Intelligence is related to ML as its subset, which is prominently used nowadays. 

Here is a catch that cosmology has sufficiently huge amount of data availability. Therefore, in this research work, we validate the ML techniques with that of theoretical model. The result analysis supports our hypothesis of making predication at par with that of the theoretical model. The ML algorithms have also been extensively explored by researchers in the field of cosmology as well. Scientists have explored these ML techniques to extract cosmological parameters \cite{ref47}, analyzing the reliability of cosmological models \cite{ref48,ref49}, and make predictions of Hubble parameters for red-shift parameters \cite{ref50,ref51,ref52}. ML techniques have been explored in different frameworks of Astronomy, cosmology, and Astrophysics \cite{ref53,ref54,ref55,ref56}. With modern ML techniques, researchers can tackle complex problems in cosmology by opening new avenues for the discovery and validation of theoretical models \cite{ref57,ref58,ref59}. From parameter estimation and model fitting to solving differential equations and classifying cosmological scenarios, ML can enhance our understanding of anisotropic universes and improve the analysis of observational data. However, besides the aforesaid references, there are several ML techniques related notable works available in the literature \cite{Arjona2002,Salti2021,Wang2021,Tilaver2021,Gomez2023,Shah2023,Lucie2024,Vilardi2025}. Interested readers may go through these works for deeper and multifarious applications in different arena of cosmological science and technology. This literature survey regarding distinguishing feature on ML techniques obviously can shed sufficient light and therefore can emphasize to utilize adequate support by references to recent and relevant studies in the field.

Therefore, inspired by the above-mentioned ML research in the cosmology domain, in the present study we have employed various ML approaches to estimate and analyze the Hubble parameter for our proposed anisotropic cosmological model of the universe and perform observational data analysis. We investigate linear regression, ANN, and Polynomial regression machine learning techniques to analyze the Bianchi I cosmological model. Bianchi's type I metric provides a valuable framework for exploring the implications of modified gravity theories in anisotropic and homogeneous cosmologies. It helps us to understand how deviations from GR can influence the universe's evolution and observable properties. The concept of anisotropy and varying scales in the Bianchi Type I metric can inspire to innovative approaches in ML, particularly in the areas of Anisotropic Data Transformation, Time-Dependent Models, and Spatial-Temporal Data Analysis. 

The paper is structured as follows: Section 1 provides an overview of the motivation behind the study, in a cosmological context and the importance of observational data analysis in constraining the parameter space of theoretical models. In Section 2, we outline the main equations of this metric. Section 3 introduces the cosmic chronometer data obtained from slowly evolving distant galaxies, which serves as the observational tool for our analysis. We discuss the significance of this data set and its relevance to constraining the parameter space of Bianchi metric. Here, we describe the statistical procedures employed in our analysis, including the $\chi^2$ minimization technique, and the Markov chain Monte Carlo method. In section 4, we discuss the methods of machine learning algorithms. In Section 5, we explain the machine learning analysis by fitting the theoretical models to the observational data whereas in Section 6 we have done a specific discussion in terms of ML view point. In the last Section 7, we conclude the research work along with future directions.

 \section{Line element and field equations} 

 In cosmology, the metric related to Bianchi space-time has been treated as a fundamental tool to explore the evolutionary properties of anisotropic universes. Basically this line element helps understanding the scenario which is beyond the curtain of the theoretically conceptualized models of isotropic universe. Therefore, open up the realm of the early universe and its anisotropic dynamical properties. However, here it will be convenient to provide a more detailed justification for the choice of the Bianchi Type I model in the context of current observational data. As such the standard cosmological model assumes the universe as of isotropic and homogeneous nature. However, the analysis of the Cosmic Microwave Background (CMB) observational data suggests that the initial universe is not with isotropic feature \cite{Spergel2003}. The Wilkinson Microwave Anisotropy Probe (WMAP) observational analysis also supports the anisotropic Behaviour of early universe \cite{ref29,Bennett2003}. Therefore, considering the CMB radiation predictions along with the WAMP confirmations, the Bianchi formulation of geometrical space seems suitable to describe anisotropy of universe. In this regard, various models based on anisotropic Bianchi formulations have been proposed to discuss cosmological events \cite{Demianski1992,Singh2016,De2022}. Recently, Nojiri et al. \cite{Nojiri2022} formulated anisotropic evolution in the context of modified gravity theory, specifically focusing on the pre-inflationary phase as well as the near-vicinity of the inflationary epochs.  
 
 Let us consider the metric in the following form for Locally Rotationally Symmetric (LRS) Bianchi type I spacetime:
\begin{equation}\label{1}
ds^2=-dt^2 + B^2 \left( dy^2+ dz^2 \right) + A^2 (dx)^2,  
\end{equation} 
where the symbols $A$ and $B$ are defined as the time-dependent metric functions. 

Now, Einstein's field equation of GR can be given by:
\begin{equation}\label{2}
R_{ij}-\frac{1}{2} g_{ij} R + \Lambda g_{ij} = - T_{ij} ,
\end{equation}
where $\Lambda$ is the erstwhile cosmological constant as was adopted by Einstein in the requirement for his static cosmological model and presently known as the dark energy component which is responsible for the late time acceleration of the universe. The tensorial energy-momentum part in its explicit format is  $T_{ij} = (p_{m}+\rho_{m})u_{i} u_{j}+p_{m} g_{ij}$, with $\rho_{m}$ and $p_{m}$ as the energy density and the matter pressure, respectively. 

Under the above GR-based framework, the field equations for the Bianchi - I universe are expressed as:
\begin{equation}\label{3}
2 \frac{\ddot{B}}{B}+ \frac{\dot{B}^2}{B^2}=-p_{m}+\Lambda,
\end{equation}

\begin{equation}\label{4}
\frac{\ddot{A}}{A}+\frac{\ddot{B}}{B}+ \frac{\dot{A} \dot{B} }{A B}=-p_{m}+\Lambda,
\end{equation}

\begin{equation}\label{5}
2\frac{\dot{A} \dot{B}}{AB}+\frac{\dot{B}^2}{B^2}=\rho_{m}+\Lambda.
\end{equation}

For our specifically proposed model, the specifications are as follows: $ a =\left(A B^{2}\right)^{1/3}$ is the average scale factor, $ V =A B^{2} $ is the volume, and $ H = \frac{\dot{a}}{a} = \frac{1}{3} \left(\frac{\dot{A}}{A}+2\frac{\dot{B}}{B}\right) $ is the Hubble parameter.

Hence, from Eqs. (\ref{3}) -- (\ref{5}), one can get \cite{ref60} 
\begin{equation}\label{6}
H^2 =\frac{1}{3}\left(\rho_{m}+\Lambda+\frac{1}{3} \frac{c_{1}^{2}}{a^6}\right),
\end{equation}
where $c_{1}$ is the constant of integration.

The energy conservation law in case of barotropic fluid, can be provided as \cite{ref38,ref61}
\begin{equation}
\frac{d}{dt} \rho_{m}+3 H \rho_{m}+3 H p_{m} =0.\nonumber
\end{equation}

We assume that the present universe is ``dust filled", i.e. $ p_{m}=0 $ so that $ \rho_{m} \propto a^{-3}$. Therefore, by the application of the relation of scale factor and redshift, one can define $1+z = \frac{a_{0}}{a}$ so that eventually it yields $ \rho_{m} = \rho_{m0} (1+z)^3 $. 

Again, for the proposed model anisotropic energy density can be derived as $\frac{1}{3} \frac{c_{1}^{2}}{a^6} =\rho_{\sigma0} (1+z)^{6} =\rho_{\sigma}$ \cite{ref62} whereas the density parameters for the given model of the universe filled with dust can be presented as $ \Omega_{\sigma} = \frac{\rho_{\sigma}}{\rho_{c}}$ and $ \Omega_{m} = \frac{\rho_{m}}{\rho_{c}}$ with the following definitions $ \rho_{c} = \frac{3 H^{2}}{8 \pi G} $ and $ 8 \pi G\approx 1$.    

Let us now rewrite Eq. (\ref{6}) in the following form:
\begin{equation}\label{7}
H^2 = H^{2}_{0}\left[ (1+z)^{3}\Omega_{m0}+\Omega_{\Lambda0}+ (1+z)^{6}\Omega_{\sigma0}\right],
\end{equation}
from which the relation between density parameters, for $ z=0 $, can be obtained as: 
\begin{equation}\label{8}
\Omega_{m0} +\Omega_{\Lambda0}+\Omega_{\sigma0} = 1. 
\end{equation}

In the above Eq. (\ref{8}), the density parameters are used to quantify the different components of the universe's density under $ z=0 $. They are defined as the ratio of the density of a specific component (e.g., matter, dark energy, anisotropy, etc.) to the critical density, which is the density required for the universe to be spatially flat. In the present Bianchi Type I cosmological model, density parameters like the matter density parameter ($\Omega_{m}$), the dark energy density parameter ($\Omega_{\Lambda}$), and the anisotropy density parameter ($\Omega_{\sigma}$) provide insights into the composition and evolution of the universe. Here $\Omega_{m}$ represents the proportion of the universe's density due to ordinary matter, while $\Omega_{\Lambda}$ represents the proportion due to dark energy. On the other hand, $\Omega_{\sigma}$ quantifies the anisotropy or lack of uniformity in the expansion of the universe. Therefore, a higher value of $\Omega_{\sigma}$ indicates a greater degree of spatial inhomogeneity and anisotropy. 

The expression of the deceleration parameter is derived for the proposed model as follows:
\begin{equation}\label{9}
q= \frac{ (1+z)^{3}\Omega_{m0}-2\Omega_{\Lambda0}+4 (1+z)^{6}\Omega_{\sigma0}}{2 \left[ (1+z)^{3} \Omega_{m0}+\Omega_{\Lambda0}+ (1+z)^{6}\Omega_{\sigma0}\right]}.
\end{equation}

It is to be noted that the Hubble parameter ($H$) and deceleration parameter ($q$) are significant physical quantities in connection to the evolutionary history of the universe. \\

\section{Datasets and cosmological constraining methodology}

\subsection{Cosmic chronometer (CC) data}

We have considered 30 $H(z)$ data points for $z$ ranging in between $0.07$ and $1.965$ calculated from cosmic chronometric technique, galaxy clusters \cite{ref32}, and differential age procedure. As usual, the Hubble constant can be realized in the form of redshift as $(1+z) H(z) = -\frac{dz}{dt}$ \cite{ref62a,ref62b,ref62c}. 

Now, the estimator $\chi^2$ is taken into consideration for the purpose of limiting the model’s parameters by comparing the model's theoretical predictions ($E_{th}$) with experimental values ($E_{obs}$) \cite{ref63,ref64,ref65,ref66,ref67,ref68,ref69}
\begin{equation}
\chi^{2}_{CC} = \sum_{i= 1}^{30} {\frac{\big[E_{th}(z_i)-E_{obs}(z_i)\big]^2}{\sigma^2_i}},
\end{equation}
where $\sigma_{i}$ is the error detected in experimental estimations of $H(z)$.

\subsection{Baryon Acoustic Oscillation (BAO) data} 

To determine the restrictions on parameters of the model, we have taken into account the Baryon Acoustic Oscillation (BAO) \cite{ref63,ref69a,ref69b} measurements dataset. Six BAO data points have been considered (Table 1). For the BAO sample, the predictions from a sample of Galaxy Surveys like SDSS DR7 and 6dF, and WiggleZ have been utilized \cite{ref63,ref69a,ref69b}. A similar explanation of the given sample can be seen in \cite{ref45,ref58}, but \cite{ref69e} provides more information on the approach used and sample to constrain the parameters.

The angular diameter distance for the sample is defined as $D_{A}=\frac{D_{L}}{(1+z)^2}$, where $D_{L}$ indicates the proper angular diameter distance \cite{ref69e}, and the dilation scale is described by $D_{V}(z)=\left[ D^{2}_{L}(z)*(1+z)^2*\frac{c \,  z}{H(z)} \right]^{1/3}$. 

\begin{table}
	\begin{center}
    \caption{Baryon Acoustic Oscillation (BAO) \cite{ref63,ref69a,ref69b} measurements dataset }
		\begin{tabular}{|c|c|c|c|c|c|c|}
			\hline 
			\multicolumn{7}{|c|}{ Values of $\varUpsilon(z)$ for different points of $z_{BAO}$}\\
			\hline
			$z_{BAO}$& $0.106$ & $0.2$	 & $0.35$  & $0.44$ & $0.6$ &$ 0.73$  \\
			\hline
			\small  $\varUpsilon(z)$ & \small$30.95 \pm 1.46$ & \small$17.55 \pm 0.60$  & \small$10.11 \pm 0.37$ & \small $8.44 \pm 0.67$ &\small $6.69 \pm 0.33$ & $ 5.45 \pm 0.31$\\
			\hline
		\end{tabular}
	\end{center}
\end{table}

In the above, $\varUpsilon(z)= D_A(z_{*})/D_V(z_{BAO})$ and $z_{*}\approx 1091$.

For limiting the parameters of the model,  the chi-square estimator for the BAO sample is described in the following form \cite{ref69e,ref69d}:
\begin{equation}
\chi^2_{BAO} = X_{BAO}^{T} C_{BAO}^{-1} X_{BAO},
\end{equation}
where 
\begin{align*}
X_{BAO} &= \begin{pmatrix}
\frac{d_A(z_{*})}{D_V(0.106)}-30.95\\           
\frac{d_A(z_{*})}{D_V(0.20)}-17.55\\
\frac{d_A(z_{*})}{D_V(0.35)}-10.11\\
\frac{d_A(z_{*})}{D_V(0.44)}-8.44\\
\frac{d_A(z_{*})}{D_V(0.60)}-6.69\\
\frac{d_A(z_{*})}{D_V(0.73)}-5.45
\end{pmatrix}
\end{align*}
and $C_{BAO}^{-1}$ is given by \cite{ref69e}
\begin{align*}\label{15}
C_{BAO}^{-1} &= \begin{pmatrix}
0.48435 & -0.101383 & -0.164945 & -0.0305703 &-0.097874 & -0.106738\\           
-0.101383 & 3.2882 &-2.45497 &-.0787898 &-0.252254 &-0.2751\\
-0.164945 &-2.454987 &9.55916 &-0.128187 &-0.410404 &-0.447574\\
-0.0305703 &-0.0787898 &-0.128187 &2.78728&-2.75632 &1.16437\\
-0.097874 &-0.252254 &-0.410404 &-2.75632 &14.9245 &-7.32441\\
-0.106738 &-0.2751 &-0.447574 &1.16437 &-7.32441 &14.5022
\end{pmatrix}.
\end{align*}

\subsection{Cosmic Microwave Background Radiation (CMBR) data}

In addition, we use the observational data of Cosmic Microwave Background Radiation (CMBR) acoustic peak \cite{ref69e,ref69c,ref69d}. The CMB measurements under consideration are based on WMAP7 observations \cite{ref69c}. In order to establish adequate constraints on DE models, the sample of CMB measurements is crucial. The position of this peak is given by $(l_a, R, z_{*})$, where $z_{*}$ representing the recombination epoch and $R$ representing the scale distance to the recombination epoch and
\begin{equation}
l_{a} = \pi \frac{D_{A}(z_{*})}{r_{s} (z_{*})}.
\end{equation}

Also, the prior distance $R$ is given by 
\begin{equation}
R = \sqrt{\Omega_{m0}} H_{0} D_{A}(z_{*}).
\end{equation}

Additionally, we use the recombination epoch fitted formula provided in \cite{ref70}. For the acoustic pick position of CMB observations, we use the WMAP dataset of correlated points \cite{ref70a}. For limiting the model parameter,  the chi-square estimator for the CMB dataset is described in the following form \cite{ref70b}
\begin{equation}
\chi^2_{CMB} = X_{CMB}^{T} C_{CMB}^{-1} X_{CMB},
\end{equation}
where 
\begin{align*}
X_{CMB} &= \begin{pmatrix}
l_{a}-302.40\\           
R-1.7264\\
z_{*}-1090.88
\end{pmatrix},
\end{align*}
and 
\begin{align*}
C_{CMB}^{-1} &= \begin{pmatrix}
3.182 & 18.253 & -1.429 \\           
18.253 & 11887.879 &-193.808 \\
-1.429 &-193.808 & 4.556 
\end{pmatrix}.
\end{align*}

Thus, the joint estimator for a combined sample of the experimental predictions including BAO, CMBR, and CC, the combined statistic measure is defined in the following manner \cite{ref66,ref67,ref68,ref70c}
\begin{equation}
\chi^2_{tot} = \chi^{2}_{CC} + \chi^2_{BAO} + \chi^2_{CMB}.
\end{equation}

The $\chi^{2}_{tot}$  statistics can be minimized to find the parameter value that best fits the combined sample of CC, BAO, and CMBR datasets.  By taking maximum likelihood approach into account, the total likelihood function $\mathcal{L}_{tot} = exp(-\chi^2_{tot}/2)$ may be calculated as the product of individual likelihood functions of each dataset expressed in the form $\mathcal{L}_{tot}=  \mathcal{L}_{BAO}*\mathcal{L}_{CMBR}* \mathcal{L}_{CC}$. The likelihood function $\mathcal{L}_{tot}(x*)$ is maximized or, alternatively $\chi^2_{tot} (x^{*})=-2 \ln \mathcal{L}_{tot} (x^{*})$ is minimized to get the most plausible values of parameters. For the set of the cosmic parameters (pointed at $x^{*}$), the $1 \sigma$ and $2 \sigma$ contours are constrained and bounded respectively by $\chi^2_{tot} (x)=\chi^2_{tot} (x^{*})+2.3$ and $\chi^2_{tot} (x)=\chi^2_{tot} (x^{*})+6.17$.  We get best-fit parameter values for the derived model by minimizing the $\chi^2$ statistics. 

\begin{figure}[H]
\centering
\includegraphics[scale=0.9]{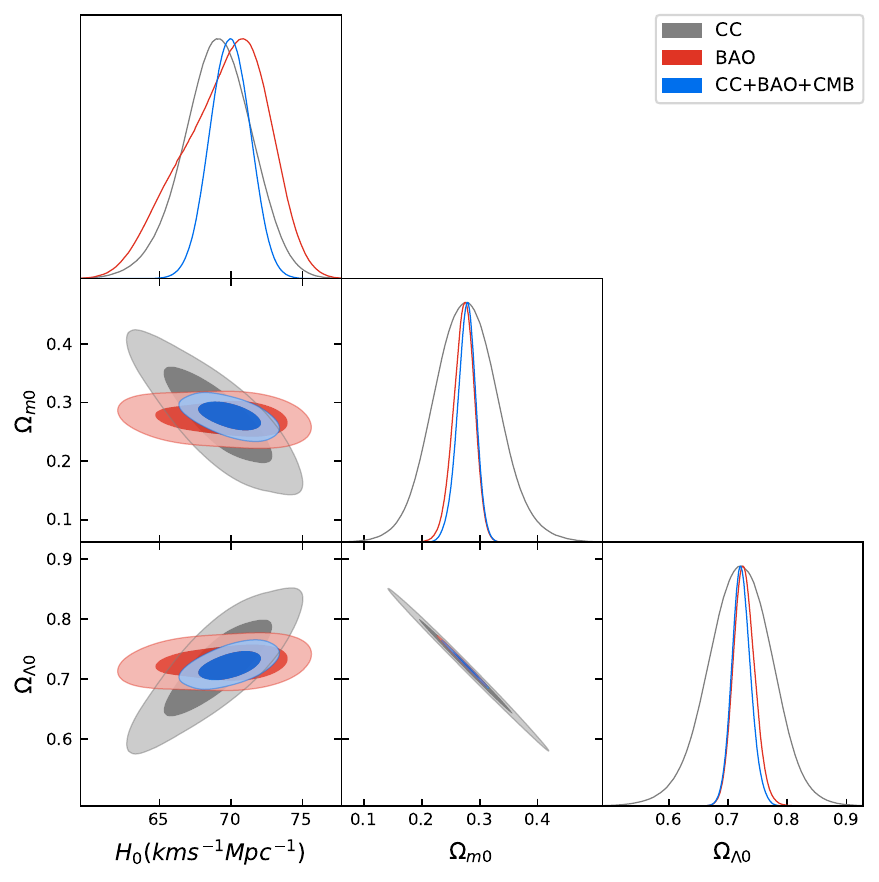}
\caption{1-dimensional marginal plots and 2-dimensional contour plot with 68$\%$ confidence level and $95 \%$  confidence level.}
\end{figure}

For the joint dataset of CC, BAO, and CMB, the parameter’s best-fit values for the derived model are estimated as $ H_0 = 69.9\pm 1.4$ km/s/Mpc, $ \Omega_{m0}=0.277^{+0.017}_{-0.015}$, $ \Omega_{\Lambda 0} = 0.722^{+0.015}_{-0.017}$,  and $\Omega_{\sigma 0} = 0.0009\pm0.0001$. Figure 1 displays the statistical results in confidence contours with $1\sigma$ and $2\sigma$ limits for the proposed model utilizing the joint dataset of BAO, CMB, and CC. The best plausible values of parameters estimated from the different dataset are summarized in Table 2. These results of the proposed model are in nice agreement with the recent observational cosmological findings~\cite{ref62a,ref71,ref72,ref73,ref76,ref77,ref78,ref79}.\\

\begin{table}[H]
	\caption{ The best-fit values model parameters for joint observational dataset }
	\begin{center}
		\begin{tabular}{|c|c|c|c|c|}
			\hline
			 Parameters &  $H_{0}$ &  $\Omega_{m0}$	& $\Omega_{\Lambda 0}$ &  $\Omega_{\sigma 0}$  \\
			\hline
			CC &  $69.1^{+2.5}_{-2.2}$ &  $0.277\pm 0.054$  &  $0.720\pm 0.053$ &  $0.003\pm0.001$  \\ 
		    \hline
			 BAO &  $69.6^{+3.5}_{-2.5}$ &  $ 0.273^{+0.019}_{-0.017}$  &  $0.726^{+0.017}_{-0.019}$ &  $0.0010\pm0.0001$  \\ 
			\hline
			 CC+BAO+CMB &  $69.9\pm 1.4$ &  $0.277^{+0.017}_{-0.015}$  &  $0.722^{+0.015}_{-0.017}$ &  $0.0009\pm0.0001$ \\
			\hline
		\end{tabular}
	\end{center}
\end{table}

\section{Machine Learning approach}

The ML techniques have revolutionized almost every field of human endeavor varying from daily life needs to science and technology \cite{ref80,ref81,ref82,ref83}. There is no exception with cosmology as well. The researchers have explored ML techniques for the analysis of cosmological events \cite{ref84,ref85}. In that line, we exploit ML techniques in our research work as well.  This section describes the ML techniques used in the present research work. These ML algorithms help us to support the results of theoretically designed mathematical model. These techniques are generally classified into supervised and unsupervised ML techniques. This work explores supervised ML algorithms as the data is labeled.  Supervised ML algorithms are used for classification and regression tasks. However, the current data contains the predicted values. The motivation for finding $H(z)$ with ML techniques is to validate the theoretical modeling results. We compare the results of the ML model with those of the theoretical model. The alignment of the ML results with the theoretical model validates that one. Therefore, we explore regression techniques where we have used three ML techniques: Linear Regression, ANN, and Polynomial Regression.

\subsection{Linear Regression (LR)}

LR is one of the basic supervised ML techniques which is quite helpful if there is a linear relationship between the input and output parameters \cite{ref86,ref87}. The predicted value $y_{pred}$ of the linear regression model is given by \cite{ref82,ref86,ref87}: 
\begin{equation}
y_{pred}= \theta_0 x_0  + \theta_1  x_1 + \theta_2 x_2......\theta_n x_n.
\end{equation}

The input values are represented using $x_0,x_1……..x_n$  and the weights are represented using $\theta_0,\theta_1……..\theta_n$. Initially, the weights are randomly assigned and the linear regression model is trained using the input available data. Further, the cost function calculates the distance between the predicted and actual values. In this manner, the model updates the weights to minimize the value of the cost function \cite{ref86,ref87,ref88}.

\subsection{Artificial Neural Network (ANN)}

Most of the ML techniques are based on deep learning where the foundation of these deep learning models is ANN. The ANN consists of three layers: One input layer, hidden layers, and one output layer. Each one of these layers consists of neurons. The number of neurons in the input layers is based on the number of inputs to the neural network. Similarly, the number of neurons in the output layers corresponds to the output of ANN. The number of the hidden layers are determined by the complexity of the ANN model.  Figure 2 depicts the architecture of ANN model. Biologically, the human brain consists of the neurons which pass the information from one neuron to another neuron. Similarly, the ANN consists of artificial neurons \cite{ref81,ref83} which assist the ML model in making predictions \cite{ref89}
\begin{figure}[H]
	\centering
	\includegraphics[scale=0.4]{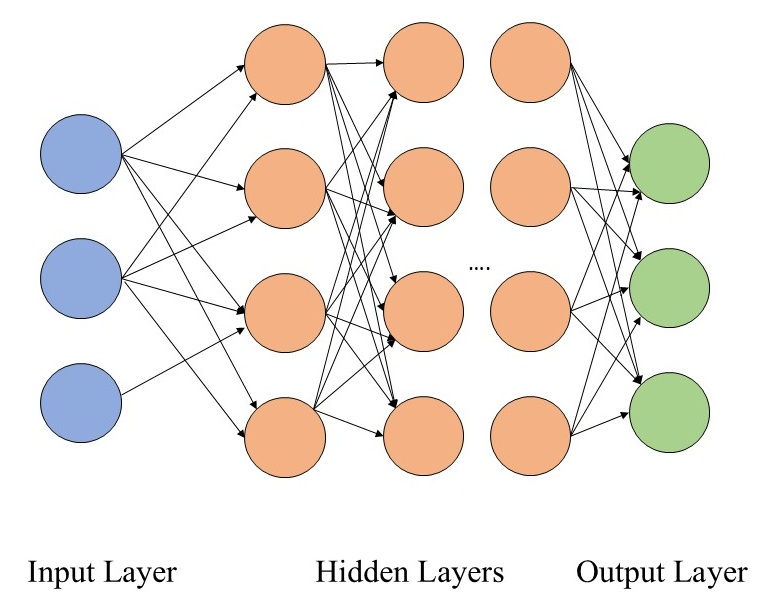}
	\caption{Layers in Artificial Neural Network (ANN).}
\end{figure}

\subsection{Polynomial Regression (PR)}

The linear regression technique is appropriate for the linear dataset. However, the relationship between input and output parameters is usually non-linear in real-world scenarios. Therefore, there is a need for an improved ML algorithm to deal with such non-linear datasets. The polynomial regression explores this non-linear relationship between the input and output data points. Using polynomial regression, we explore the dependency of output, $y$, not only on single input variable $x$, but also on its powers, i.e., $x^n$ where the parameter $n$ represents the degree of polynomial regression. In the current research work, we perform simulation using the degree of polynomial regression as 2. This type of regression (i.e. degree =2) is also termed as the quadratic polynomial regression \cite{ref82}.  

Another way to implement the polynomial regression is to change the dimension of the input features and then apply the linear regression. For example, if there are two input features [a, b], then the polynomial features are as follows: $[1, a, b, ab, a^2, b^2]$ \cite{ref82,ref90}. After that we apply linear regression on the obtained features.

 \begin{table}[H]
 	\caption{The comparative analysis of $H(z)$ corresponding to observatory, theoretical, and the machine learning predictions.}
 	\begin{center}
 		\begin{tabular}{|c|c|c|c|c|c|c|c|c|c|}
 			\hline
 			$z$& $H_{obs}$&  $H_{theo}$&	$H_{LR}$&	$H_{ANN}$& $H_{Poly_{LR}}$& $\alpha_{theo}$ & $\alpha_{LR}$& $\alpha_{ANN}$& $\alpha_{Poly_{LR}}$\\
 			\hline
 			0.07& 69&	72.121&	66.0021&	66.3296&	71.6117&	0.0452&	0.0434&	0.0387&	0.0379\\
 			\hline
 			0.09&	69&	72.779&	67.2701&	67.5871&	72.3592&	0.0548&	0.025&1	0.0205&	0.0487	\\
 			\hline
 			0.12&	68.6&	73.8&	69.172&	69.4735&	73.5042&	0.0758&	0.0083&	0.0127&	0.0715	\\
 			\hline
 			0.17&	83&	75.593&	72.3419&	72.6174&	75.4763&	0.0892&	0.1284&	0.1251&	0.0906
 			\\
 			\hline
 			0.18&	75&	75.928&	72.9759&	73.2462&	75.8802&	0.0124&	0.027&	0.0234&	0.0117
 			\\
 			\hline
 			0.2&	75&	76.697&	74.2438&	74.5037&	76.6977&	0.0226&	0.0101&	0.0066&	0.0226
 			\\
 			\hline
 			0.2&	72.9&	76.723&	74.2438&	74.5037&	76.6977&	0.0524&	0.0184&	0.022&	0.0521\\
 			\hline
 			0.27&	77&	79.518&	78.6817&	78.9052&	79.6591&	0.0327&	0.0218&	0.0247&	0.0345
 			\\
 			\hline
 			0.28&	88.8&	79.934&	79.3157&	79.534&	80.0949&	0.0998&	0.1068&	0.1043&	0.098
 			\\
 			\hline
 			0.35&	83&	83.064&	83.7535&	83.9355&	83.2345&	0.0008&	0.0091&	0.0113&	0.0028
 			\\
 			\hline
 			0.38&	83&	84.35&	85.6554&	85.8218&	84.6277&	0.0163&	0.032&	0.034&	0.0196
 			\\
 			\hline
 			0.4&	95&	85.272&	86.9234&	87.0794&	85.5725&	0.1024&	0.085&	0.0834&	0.0992
 			\\
 			\hline
 			0.4&	77&	85.291&	86.9234&	87.0794&	85.5725&	0.1077&	0.1289&	0.1309&	0.1113
 			\\
 			\hline
 			0.42&	87.1&	86.446&	88.1913&	88.337&	86.53&	0.0075&	0.0125&	0.0142&	0.0065 \\
 			\hline
 			0.45&	92.8&	87.66&	90.0933&	90.2233&	87.9901&	0.0554&	0.0292&	0.0278&	0.0518\\
 			\hline
 			0.48&	80.9&	89.079&	91.9952&	92.1096&	89.4788&	0.1011&	0.1371&	0.1386&	0.106\\
 			\hline
 			0.48&	97&	89.165&	91.9952&	92.1096&	89.4788&	0.0808&	0.0516&	0.0504&	0.0775
 			\\
 			\hline
 			0.59&	104&	95.095&	98.9689&	99.0263&	95.1826&	0.0856&	0.0484&	0.0478&	0.0848\\
 			\hline
 			0.68&	92&	99.988&	104.6747&	104.6853&	100.1357&	0.0868&	0.1378&	0.1379&	0.0884
 			\\
 			\hline
 			0.78&	105&	106.006&	111.0145&	110.9731&	105.9415&	0.0096&	0.0573&	0.0569&	0.009\\
 			\hline
 			0.88&	125&	111.916&	117.3543&	117.2609&	112.0656&	0.1047&	0.0612&	0.0619&	0.1035	\\
 			\hline
 			0.88&	90&	112.239&	117.3543&	117.2609&	112.0656&	0.2471&	0.3039&	0.3029&	0.2452\\
 			\hline
 			0.9&	117&	113.536&	118.6222&	118.5185&	113.3286&	0.0296&	0.0139&	0.013&	0.0314\\
 			\hline
 			1.04&	154&	122.756&	127.4979&	127.3214&	122.526&	0.2029&	0.1721&	0.1732&	0.2044\\
 			\hline
 			1.3&	168&	141.964&	143.9813&	143.6698&	141.2617&	0.155&	0.143&	0.1448&	0.1592\\
 			\hline
 			1.36&	160&	146.837&	147.7851&	147.4425&	145.8909&	0.0823&	0.0763&	0.0785&	0.0882\\
 			\hline
 			1.43&	177&	152.128&	152.223&	151.8439&	151.4363&	0.1405&	0.14&	0.1421&	0.1444\\
 			\hline
 			1.53&	140&	160.226&	158.5627&	158.1318&	159.6289&	0.1445&	0.1326&	0.1295&	0.1402\\
 			\hline
 			1.75&	202&	178.848&	172.5102&	171.965&	178.7728&	0.1146&	0.146&	0.1487&	0.115\\
 			\hline
 			1.96&	186.5&	198.046&	185.8237&	185.1694&	198.4833&	0.0619&	0.0036&	0.0071&	0.0643\\
 			\hline
 		\end{tabular}
 	\end{center}\label{table:kysymys}
 \end{table}  

\section{Analysis of the ML methodology}

\begin{figure}[H]
\centering
\includegraphics[scale = 0.5]{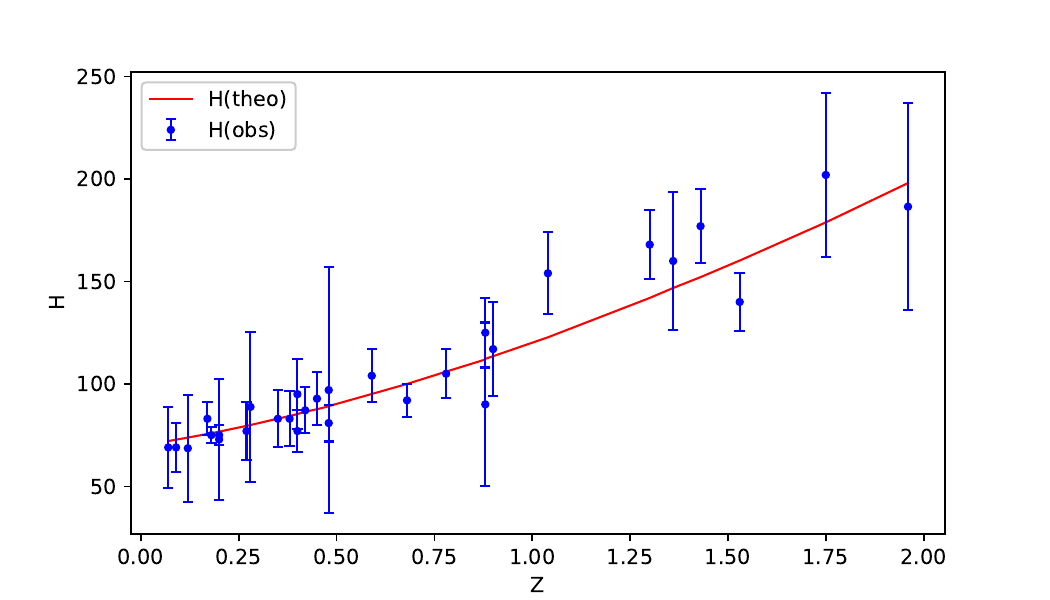}
\caption{The figure depicts observed ($H_{theo}$) and theoretical ($H_{obs}$) values corresponding to different $z$ parameters.}\label{obs_theo}
\end{figure}

In the current research work, we use the ML techniques, i.e., linear regression, ANN and polynomial regression for the analysis of Bianchi I model. The ML analysis involves two steps: first training and then testing of the model. We consider 30 points of dataset for the computation, which is further divided as $67\%$ of the data as training, and the rest $33\%$ as testing data. After splitting the data, we apply ML techniques, i.e., Linear regression, ANN, and Polynomial regression. Figure \ref{obs_theo} represents the graph between red shift parameter ($z$), corresponding to theoretical ($H_{theo}$) and observed ($H_{obs}$) values. Further, Fig. \ref{obs_theo}  depicts that the theoretical values are aligned with the observed values. 
	
\subsection{LR analysis }

\begin{figure}[H]
\centering
\includegraphics[scale = 0.5]{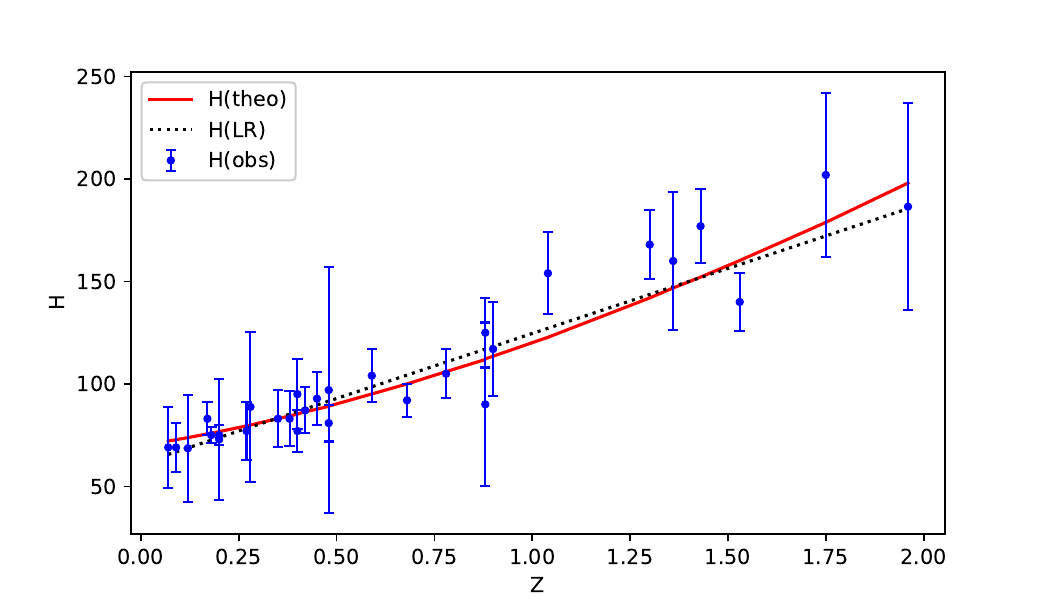}
\caption{Predictions for linear regression model corresponding to theoretical and observed values.}\label{LR}
\end{figure}

In this Section, we perform the graphical and quantitative analysis of the linear regression technique with respect to theoretical model and observed data. Figure \ref{LR} demonstrates the comparison of values of $H(z)$ predicted by the linear regression algorithm with that of observed output $H(obs)$, and theoretical model, i.e., Bianchi model. This figure exhibits that the performance of the linear regression model is quite satisfactory. The predictions made by the linear regression model is comparable to theoretical model. Further, in order to get the quantitative analysis of the linear regression, we compute the deviation for both the theoretical and linear regression model. The deviation of the theoretical value to the observed value is calculated by using Eq. \cite{ref48}: 
\begin{equation}
\alpha_{theo}=|\frac{H_{theo}}{H(obs)}-1|,
\end{equation}

\begin{equation}
\alpha_{LR}=|\frac{H_{LR}}{H(obs)}-1|.
\end{equation}

The columns 2,~3 and 4 of the Table \ref{table:kysymys} demonstrate the observed values, theoretical values and LR output values. Further, the columns 7 and 8 of the Table \ref{table:kysymys}, indicate the theoretical deviation ($\alpha_{theo}$) and linear regression deviation ($\alpha_{LR}$). In order to interpret  these values,  we take the mean of these deviations, which is given as: 
\begin{equation}
(\bar{\alpha}_{theo},\bar{\alpha}_{LR}) =(0.08073,0.07703).
\end{equation}

The difference between the values of the `mean alpha deviation' for theoretical and linear regression models is much less, which indicates that the linear regression analysis supports the results of the theoretical model.  

\subsection{ANN analysis}
\begin{figure}[H]
\centering
\includegraphics[scale = 0.5]{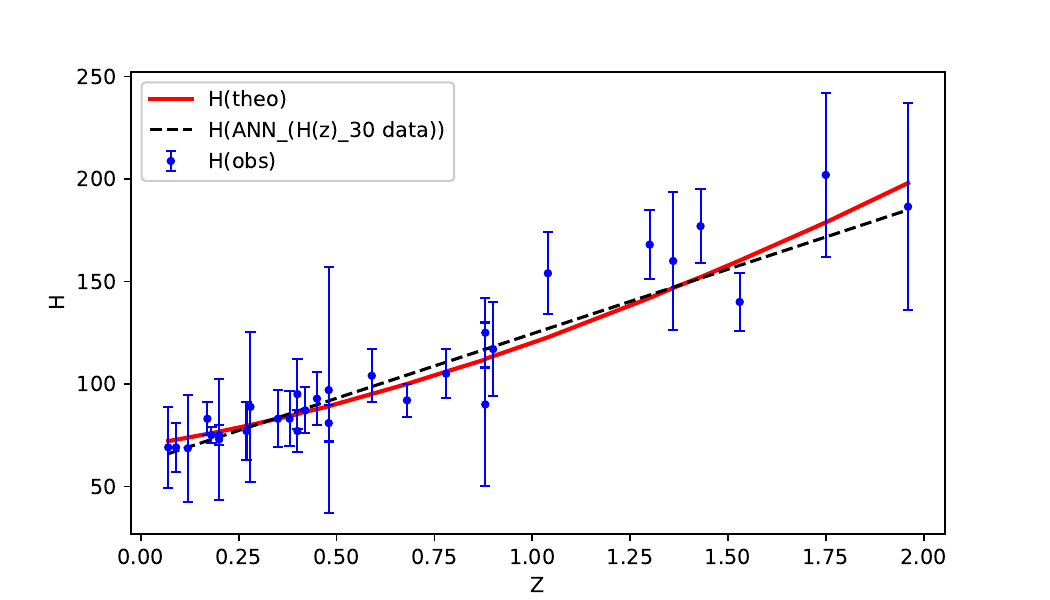}
\caption{Graphical Analysis of predictions using ANN with its theoretical and observed counter part.}\label{ANN}
\end{figure}

Similar to LR analysis, we utilize graphical and quantitative analysis to understand the effect of another ML technique, i.e., ANN. The Fig. \ref{ANN} depicts the predictions of ANN corresponds to observed data and theoretical model output. This figure indicates that the ANN performs inline with that of the theoretical model output. Further, we utilize `alpha deviation' to compute the quantitative analysis of the results. The value of deviation is given by the following equation:
\begin{equation}
\alpha_{ANN}=|\frac{H_{ANN}}{H(obs)}-1|.
\end{equation}

The columns 5 and 9  of the Table \ref{table:kysymys} indicate the output of ANN model and  the ANN deviation, respectively. 

In order to perform statistical analysis, we take mean value of these data, which is given as follows:
\begin{equation}
(\bar{\alpha}_{theo},~\bar{\alpha}_{ANN}) =(0.08073,~0.07710).
\end{equation}

In the analysis, the learning rate of the neural network is 0.001, and ‘adam’ as the optimizer of the neural network. Further, we have used ‘relu’ activation function in the present research work. The quantitative analysis supports the results of graphical representation. This indicates that the results of theoretical model and  that of the ANN are quite similar.

\subsection{PR analysis}

\begin{figure}[H]
\centering
\includegraphics[scale = 0.5]{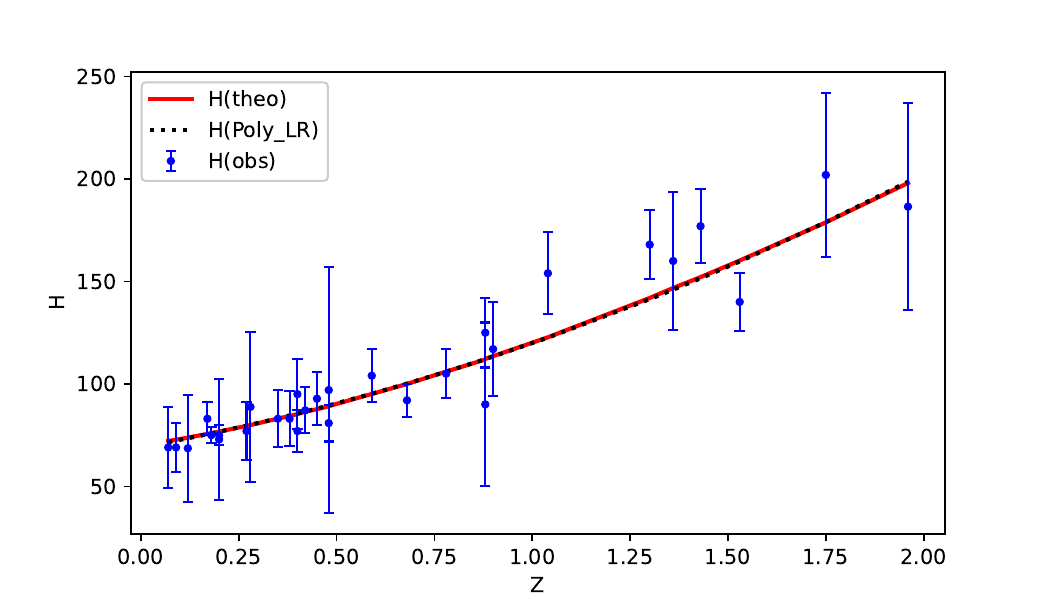}
\caption{Comparative analysis of predictions for Polynomial regression model with that of theoretical and observed values.}\label{POLY}
\end{figure}

Apart from linear regression and  ANN, we perform analysis by Polynomial regression using both the graphical and quantitative analysis. The Fig. \ref{POLY} displays the results of polynomial analysis.  This figure illustrates that the results of the polynomial regression coincides the output of the theoretical values. Further, we use quantitative data to perform the analysis. 
	
The deviation in the alpha value for polynomial regression is given by using the following formula:
\begin{equation}
\alpha_{Poly_LR}=|\frac{H_{Poly_LR}}{H(obs)}-1|.
\end{equation}

The columns 6 and 10 of the Table \ref{table:kysymys}  depict  the polynomial regression output and deviation of the polynomial regression, respectively. The average value of the Polynomial regression deviation is given as:	
\begin{equation}
	(\bar{\alpha}_{theo},~\bar{\alpha}_{Poly_LR}) =(0.08073,~0.08068).
\end{equation}

The mean polynomial deviation indicates that its value is nearly equal to the theoretical model output. Thus, the polynomial regression model reinforces the theoretical model.
	
\begin{figure}[H]
		\centering
		(a)\includegraphics[width=6cm,height=4.5cm,angle=0]{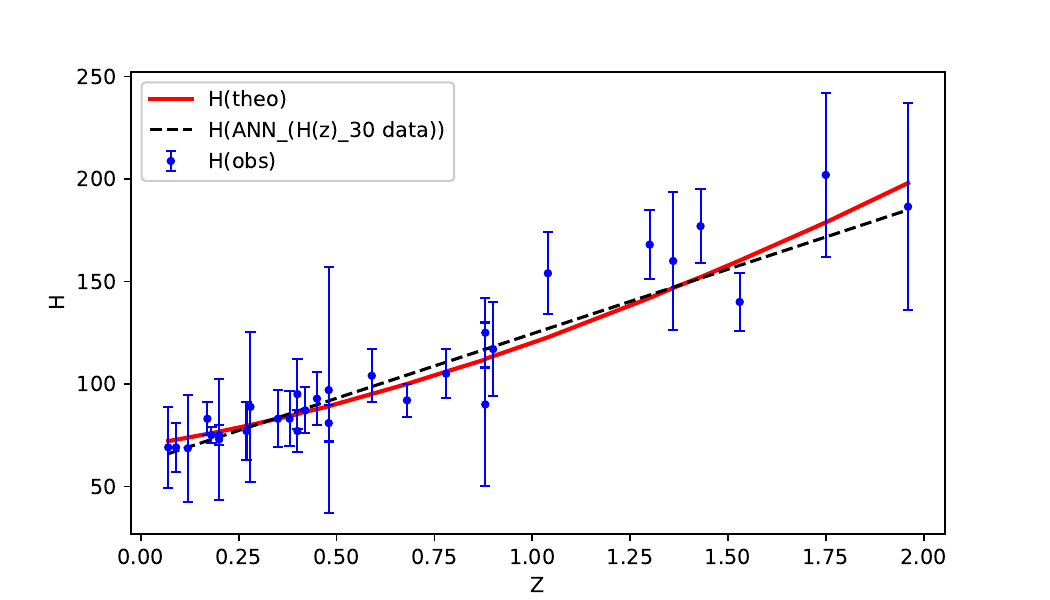}
		(b) \includegraphics[width=6cm,height=4.5cm,angle=0]{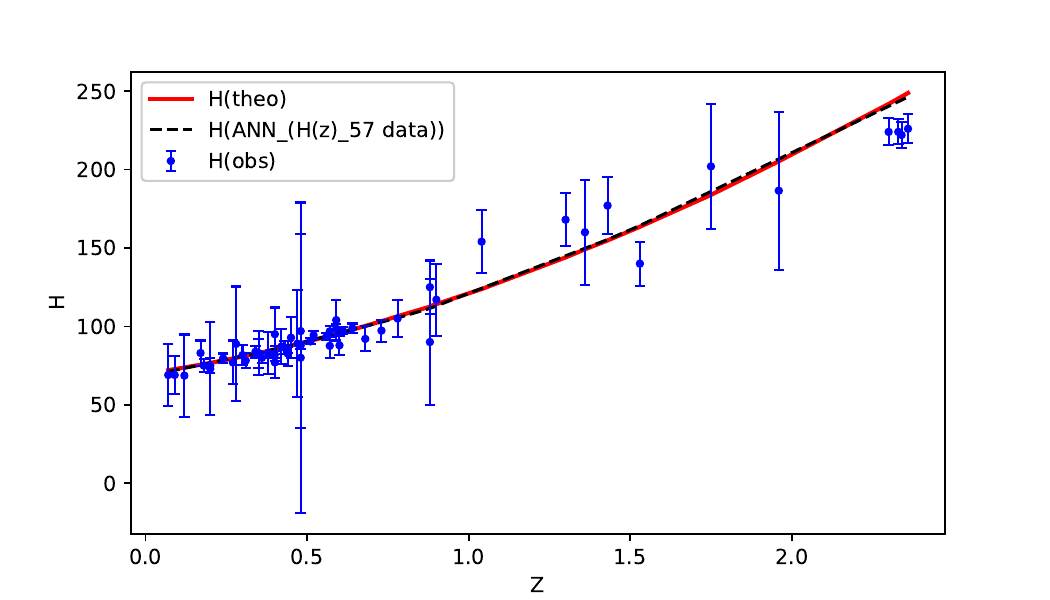}
		\caption{(a) ANN analysis with 30 $H(z)$ points, (b) ANN analysis with 57 $H(z)$ points.}\label{ANN_30}
\end{figure}

\begin{table}[H] 
		\begin{center}
		\caption{For this analysis, we split the data into training and testing data in the ratio 80\%  and 20\%, respectively where the table depicts the results obtained on the testing data.} \label{tab:performance_metrics} 
				\begin{tabular}{|c|c|c|c|c|c|c|}
                \hline
                \multirow{2}{*}{\tiny Performance matrices} & \multicolumn{3}{c|}{\tiny \textbf{Data 30}} &  \multicolumn{3}{c|}{\tiny \textbf{Data 57}} \\ \cline{2-7} 
				& \tiny \textbf{LR} & \tiny \textbf{ANN} & \tiny \textbf{Polynomial regression} & \tiny \textbf{LR} & \tiny \textbf{ANN} &  \tiny \textbf{Polynomial regression} \\
			\hline
				\tiny Mean absolute error  & \tiny 2.63897 & \tiny 2.74179 & \tiny	0.32359	&  \tiny 6.49796 &	\tiny 0.58316 &	\tiny 0.24508 \\
			\hline
			\tiny root mean square error & \tiny 3.52990 & \tiny 3.62650 & \tiny 0.38413 &	\tiny 8.55656 &	\tiny 0.99206 &	\tiny 0.29676 \\
			\hline
			\tiny $R^2$ & \tiny 0.99213 & \tiny 0.99170	& \tiny 0.99990	& \tiny 0.97984	& \tiny 0.99972	& \tiny 0.99998 \\
			\hline 
		\end{tabular}
        \end{center}
	\end{table}

\section{Detailed discussion on the results in terms of Machine Learning point of view }
\begin{enumerate}

\item[(1)] In our investigation, we introduce the ``alpha deviation" metric to quantify the discrepancy between the predicted and observed Hubble parameter, but what we feel that the analysis would benefit from including standard ML regression performance metrics such as Mean Absolute Error, Root Mean Squared Error, and the coefficient of determination ($R^2$). Hence, we have also incorporated the following performance parameters where Table 4 exhibits these results in details for two different datasets, i.e., 30 and 57 data points (also vide Fig. 7). Interestingly, the analysis of these results is in line with that of alpha deviation, especially for the data points 57 the exhibited result is much appreciable. 

\item[(2)] The study employs overall two datasets: 30 and 57 data points where the former has been divided into 67\% for training and 33\% for testing. While for a larger number of data points, i.e. 57 in the latter case, we have taken 80\% and 20\%, respectively. We compare the data for both 30 and 57 data points using Table 4. We also include Fig. 7 depicting the results with both 30 and 57 points. This analysis illustrates that quantitatively as well as qualitatively appreciable differences can be observed in the performance with increased data points.

\item[(3)] To prevent overfitting of the ML model, we increase the number of points in data points to 57 from the initial consideration of 30 data points. Further, we also increase the portion of the training set from 67\% to 80\% to eliminate the overfitting in the model. In order to check the generalizability of the proposed algorithm, we split the data between training and testing. After that, we train the model on 80\% of the data, and test it on the rest of 20\% data. This test data is completely unseen by the ML model. The Table 4 depicts the results which validate that ML models are capable of making the desired predictions.\\

	\begin{table}[h!]
	 \centering
	\caption{The table demonstrates the results after applying the machine learning algorithms on the observatory data, which consists of 57 data points. We divide the existing dataset into train and test data in the ratio of 80\% and 20\% respectively and the table illustrates the results on test data.}
	\label{Table_obs}
    \vspace{0.50cm}
	\begin{tabular}{|c|c|c|c|}
		\hline
		\tiny \textbf{Performance Matrices} & \tiny \textbf{Linear Regression} & \tiny \textbf{ANN} & \tiny \textbf{Polynomial Regression}  \\
		\hline
		\tiny \textbf{Mean Absolute Error} & 4.41818 & 4.25516  & 4.45609  \\
		\hline
		\tiny \textbf{Root Mean Square Error} & 6.13486 & 5.25461  & 5.99684  \\
		\hline
		\tiny \textbf{{$R^2$}} & 0.98637 & 0.99000  & 0.98697  \\
		\hline
	\end{tabular}
	\end{table}

	\begin{table}[h!]
     \caption{The table demonstrates the results after applying the machine learning algorithms on the observatory data for cross-validation.}
	\label{Table_cross_validation}
    \vspace{0.50cm}
\begin{tabular}{|c|c|c|c|}
		\hline
		\tiny \textbf{Performance Matrices} & \tiny \textbf{Linear Regression} & \tiny \textbf{ANN} & \tiny \textbf{Polynomial Regression}  \\
		\hline
		\tiny \textbf{Mean Absolute Error} & 7.7299 ± 3.8837 & 7.1573 ± 3.3401  & 7.07311 ± 3.6545  \\
		\hline
		\tiny \textbf{Root Mean Square Error} & 9.6176 ± 4.7425 & 9.4353 ± 4.5022  & 9.37922 ± 5.0727 \\
		\hline
		\tiny \textbf{$R^2$} & 0.84916 ± 0.18827 & 0.84361± 0.18654  & 0.83338 ± 0.23859  \\
		\hline
	\end{tabular}
	\end{table}

	\begin{figure}[h!] 
		\centering 
		\includegraphics[width=0.5\textwidth]{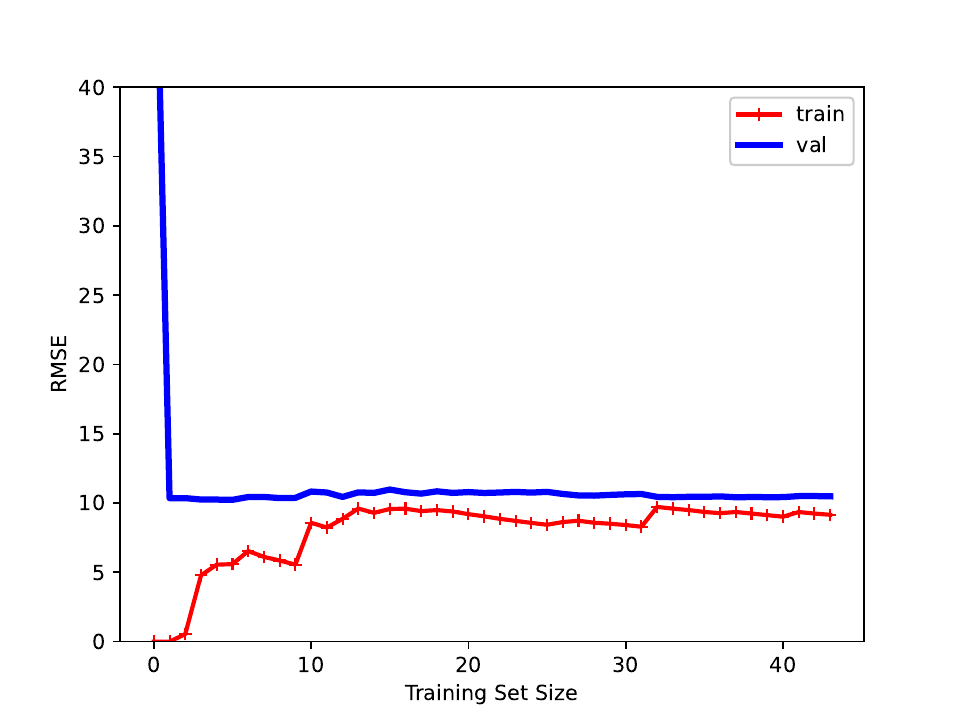} 
		\caption{Learning Curve for linear regression model. } 
		\label{fig:learning_curve_LR} 
	\end{figure}

	\begin{figure}[h!] 
		\centering
		\includegraphics[width=0.5\textwidth]{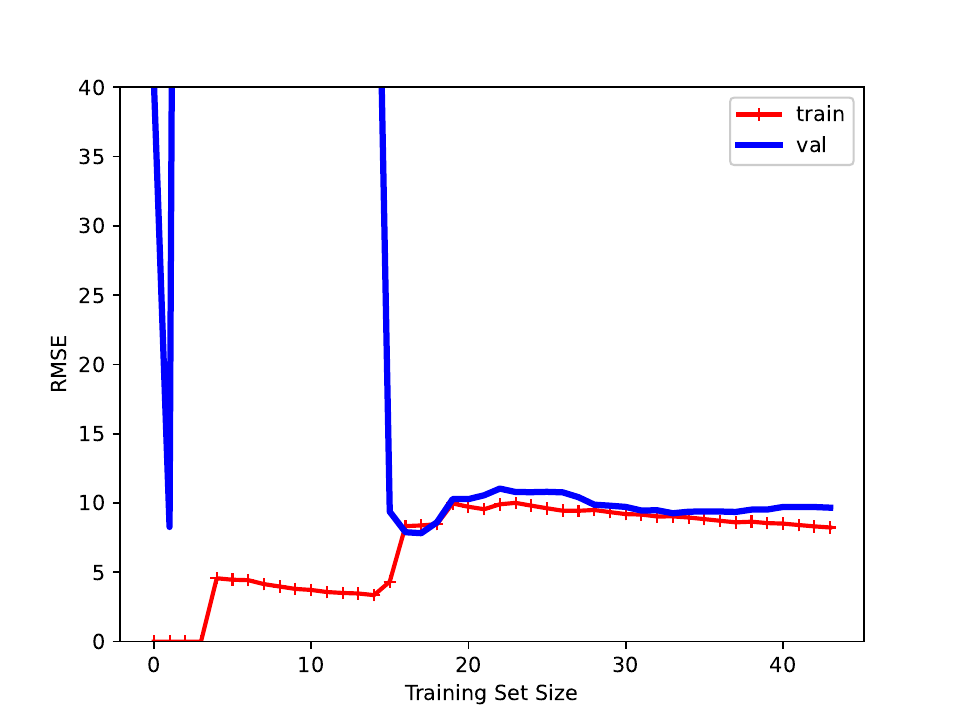} 
		\caption{Learning curve for polynomial regression. } 
		\label{fig:learning_curve_poly} 
	\end{figure}

     \begin{figure}[h!] 
        	\centering 
        	\includegraphics[width=0.5\textwidth]{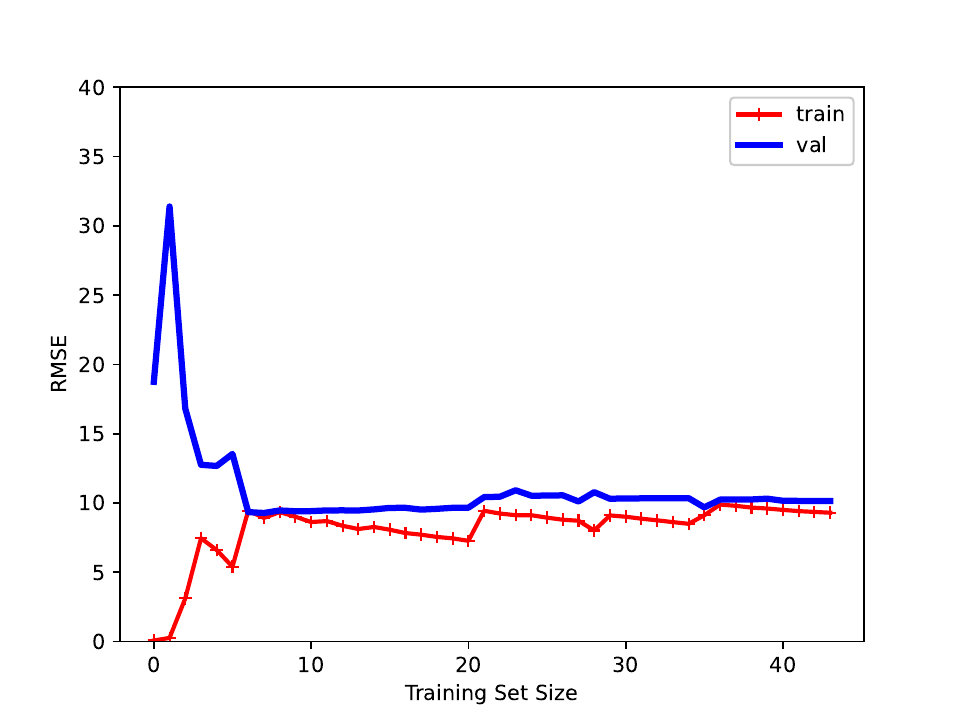} 
        	\caption{Learning curve for ANN technique on 57 data points. } 
        	\label{fig:learning_curve_ann} 
     \end{figure}

\item[(4)] To provide model independent insights, we analyze the ML models on the observatory data consisting of 57 data points. We split this data as 80\% training, and 20\%  as test data. Further, we train the machine learning models on train data and evaluate the model on the test data. Table 5 illustrates test data results. The performance matrices indicate that the ML models are capable of providing the predication independent of mathematical model.  \\	
\item[(5)] We use cross-validation technique to generalize the ML models. In current research work, we use 10 fold cross-validation, which implies that instead of single train-test split, we train the model several times by using 9 folds for the training the model and 1 fold for validating the model. In this manner, we take the 10 different combinations of train and test data to train and evaluate the ML models in order to get generalize results. After that we take the mean and standard deviations of these results, which are depicted by the Table 6. These values demonstrate the possible chances of overfitting for the ML models. \\
\item[(6)] In order to further analyze the generalization ability of ML models, we plot learning curve. Figure \ref{fig:learning_curve_LR} depicts the learning curve for the linear regression ML model. Initially, the RMSE is very small for the training and high for validation data because the model is trained on the smaller data. But as the training set size increases the validation and  training curve reach near to each other. But the root mean square error is still very high. This curve indicates the model is underfitted. Therefore, we explore complex model compared to linear regression technique, i.e., polynomial regression and ANN. Figures \ref{fig:learning_curve_poly} and \ref{fig:learning_curve_ann} demonstrate lower value of RMSE on training data and higher value on validation data to indicate the overfitting of the ML models. Therefore, we explore the regularization techniques.      \\		
\item[(7)] To further generalize the ML model, we perform the early stopping method. In this technique, as the model is trained on the data, the RMSE starts reducing for both the training and validation datasets with an increase in the epoch. But after a certain point, the validation error does not improve, and the model starts overfitting. Therefore, we explore the early stopping method. In this technique, the training of the model is stopped when the validation error is minimum. Because after that, the model is overfitted.  We have used 1000 epochs for the analysis. Stochastic Gradient Descent linear model and L2 regularization is used to eliminate the overfitting. In this analysis the best epoch is 190, and the validation error (root mean square) is 3.93948 for the polynomial regression. 
\end{enumerate}

\section{Conclusion}

In the current study, we present the observational data constraints on the parameters space for an anisotropic cosmological model of Bianchi I spacetime under GR framework. For the analysis, we consider observational datasets of CC, BAO, and CMBR peak parameters. The MCMC technique is utilized to constrain the best-fit values of the model parameters. For this purpose, we have used the publicly available Python code from CosmoMC and developed the contour plots with different constraint limits. For the joint dataset of CC, BAO, and CMBR, the parameter’s best-fit values for the derived model are estimated as $ H_0 = 69.9\pm 1.4$ km/s/Mpc, $ \Omega_{m0}=0.277^{+0.017}_{-0.015}$, $ \Omega_{\Lambda 0} = 0.722^{+0.015}_{-0.017}$ and $\Omega_{\sigma 0} = 0.0009\pm0.0001$. 

The summary of the best-fit values for various datasets can explicitly be noted in Table 2. These results for the proposed model are in nice agreement with the recent observational cosmological findings \cite{ref54,ref55,ref56,ref57,ref58,ref59}. It is now-a-days customary to note that ML techniques have been explored in the various domains of the human effort varying from space to everyday human needs. In this research article, we explore ML techniques for the analysis of the Bianchi I model. We utilize state-of-art ML techniques like linear regression, ANN and polynomial regression. The alpha deviation parameter is used to analyze the performance of the ML techniques. In the present analysis, the mean alpha deviation for theoretical and linear regression models is estimated as (0.08073,~0.07703), for theoretical and ANN regression models as (0.08073,~0.07710) while theoretical and polynomial regression models as (0.08073,~0.08068), respectively.  

Further, we perform graphical and quantitative analyses to evaluate the performance of the ML techniques. These analyses indicate that the results of the ML models are aligned with that of the theoretical model of Bianchi I spacetime. Within the applied ML techniques, the polynomial regression technique surpasses other ML techniques, like linear regression and ANN. The alignment of the predictions with the theoretical model of Bianchi I and observed values indicate that the ML algorithms perform satisfactorily for the given dataset \cite{ref48,ref52}. 

Therefore, future work involves the exploration of these ML techniques with a greater number of data points. It is to be noted ML algorithms interpret the data and successfully extract the Hubble parameter without involving the theoretical model. The predictions in the present study for all ML approaches are specifically based on the datasets. 

Basically, in this research work, we validate the ML techniques of computer science with the theoretical modeling of cosmology which reveals robustness of the methodology in the cosmic science.

\section*{CRediT authorship contribution statement}
{\bf Vinod Kumar Bhardwaj} : Conceptualization, Ideas, Formulation, Methodology. {\bf Manish Kalra}: Conceptualization, Analysis of figures and review of literature. {\bf Priyanka Garg}: Formal analysis, Writing – original, review \& editing. {\bf Saibal Ray}: Formal analysis, Supervision, Final writing \& editing.

\section*{Declaration of competing interest}
The authors declare that they have no known competing financial interests or personal relationships that could have appeared to influence the work reported in this paper.

\section*{Data Availability Statement}
All data generated or analyzed during this study are included in this published article (and its supplementary information files).

\section*{Acknowledgements}
SR would like to acknowledge Inter-University Centre for Astronomy and Astrophysics (IUCAA), Pune, India for providing them Visiting Associateship under which a part of this work was carried out and also thankful to the facilities under ICARD, Pune at CCASS, GLA University, Mathura. We all are thankful to the anonymous referee for providing pertinent suggestions which have helped us to upgrade the quality of the manuscript substantially. \\

\end{document}